\documentclass[final,5p,times,twocolumn]{elsarticle}

\usepackage{amssymb}

\journal{Physica C}

\begin{document}

\begin{frontmatter}



\title{Critical temperatures and critical currents of wide and narrow quasi-one-dimensional superconducting aluminum structures in zero magnetic field}


\author{V.~I.~Kuznetsov \corref{cor1}}
\ead{kuznetcvova@mail.ru}
\cortext[cor1]{corresponding author}
\author{O.~V.~Trofimov}
\address{Institute of Microelectronics Technology and High
Purity Materials, Russian Academy of Sciences, Chernogolovka,
Moscow Region 142432, Russia}

\begin{abstract}
We measured the critical temperatures and critical switching and
retrapping currents of wide and narrow thin-film
quasi-one-dimensional superconducting aluminum structures of the
same thickness in zero magnetic field. For the first time, we
found that the narrower the structure, the lower the critical
temperature and critical current density in the structure.
Probably, the influence of depairing centers that are on dirty
longitudinal boundaries of the structure, is the stronger than the
narrower the structure. It is found for the first time that, in
most cases, the temperature-dependent switching critical current
in both structures is approximated by two functions. At
temperatures below the temperature corresponding to the bottom of
the resistive N-S transition of structures, the switching critical
current is described by the Kupriyanov-Lukichev theory. At
temperatures close to the top of the N-S transition, the switching
current is linear with temperature and coincides with the critical
Josephson current. At these temperatures, Josephson SNS junctions
are formed in structures.
\end{abstract}




\end{frontmatter}


\section{Introduction}

Superconducting critical current is the most important
characteristic of superconducting and hybrid devices. The critical
current of superconducting quasi-one-dimensional (i.e., having
transverse dimensions less than twice the superconducting
temperature-dependent coherence length $2\xi(T)$) of wires and
various structures with weak link has been widely studied
\cite{barone, likharev, golubov, arut}.

In nanodevices, the critical current can have specific features
due to thermally activated and quantum fluctuations of the
superconducting order parameter, nonlocal and other effects.
"Nonlocality" is the mutual influence of electronic transport in
different parts of the structure.

In superconducting structures, nonlocal effects appear at a
shorter length, close to the coherence length $\xi(T)$ and at a
larger length, close to the relaxation length of the quasiparticle
charge imbalance $\lambda_{Q}(T,B)$, which depends on temperature
and magnetic field \cite{schmidt}.

It is surprising that the critical current and nonlocal
(equilibrium and nonequilibrium) phenomena in a long
quasi-one-dimensional wire with a total length several times
greater than twice the length of the quasiparticle imbalance
$2\lambda_{Q}(T,B)$ and having a transverse narrowing with a
length in the range from $\xi(T)$ to $\lambda_{Q}(T,B)$ are
practically not studied. Also, the critical current and nonlocal
effects in a long quasi-one-dimensional wire, consisting of wire
segments of different widths, have not been studied. It should be
expected that the critical current as a function of temperature
and magnetic field, measured on a short section of the wire, will
depend on the values of the superconducting order parameter in
other parts of the current-carrying wire of variable width and in
potential quasi-one-dimensional wires.

Quasi-one-dimensional aluminum wires are part of many
superconducting nanodevices. It has long been known \cite{chubov,
borisenko} that the thinner the aluminum film, the higher the
critical temperature of this film. It is also considered an
established fact that the smaller the diameter
quasi-one-dimensional superconducting aluminum wire, the higher
the critical temperature of the wire. It should be expected that
the narrower the quasi-one-dimensional thin-film aluminum wire,
the higher the critical temperature. Analyzing new effects in
superconducting quasi-one-dimensional aluminum wires of different
widths and structures consisting of such wires
\cite{kuznjetplet16, kuznphysica20, dubjetplet03, karpii07,
nikulov07, kuznprb08, kuznphysica13, kuznjetplet19}, we assumed
that another statement is valid for the structures of these works,
namely, that the smaller the wire width, the lower the critical
temperature of the wire.

This assumption that the narrow parts of the structure have a
critical temperature lower than the critical temperature of a wide
part of the structure is a necessary condition for the appearance
of negative nonlocal and local voltages (resistances) in a
superconducting quasi-one-dimensional dc-biased aluminum structure
at temperatures close to the critical temperature
\cite{kuznphysica20}.

We believe that the assumption of different critical temperatures
of the wide and narrow parts of the circularly-asymmetric aluminum
ring will help to remove the long-term intractable challenge,
which consists in the unexplained mysterious phase shift in the
magnetic field of critical currents of opposite polarities in
circularly-asymmetric aluminum structures permeated with a
magnetic flux \cite{karpii07, nikulov07}.

This phase shift of the critical currents is the reason for the
appearance of a rectified time-averaged direct voltage
$V_{rec}(\Phi/\Phi_{0})$, oscillating with a period equal to the
superconducting quantum of the magnetic flux $\Phi_{0}$, in a
circularly-asymmetric superconducting aluminum ring (a structure
of such rings in a series), permeated with the flux $\Phi$, when
an alternating current (with a zero dc component) with an
amplitude close to the critical current is passed through this
ring (structure) at temperatures slightly below the critical
temperature $T_{c}$ \cite{dubjetplet03, karpii07, nikulov07,
kuznprb08, kuznphysica13}. The circular asymmetry of a ring with a
constant inner radius $r_{i}$ is due to the fact that the ratio of
the widths of the wide and narrow semirings was usually
$w_{w}/w_{n}=2$. Such a ring can be used in an asymmetric dc
micro-SQUID \cite{barone, clarke}. Moreover, such a ring and a
structure of such rings in series can operate as a highly
efficient magnetic field-dependent ac voltage rectifier and as a
highly sensitive detector of nonequilibrium electromagnetic noise
\cite{dubjetplet03, karpii07, nikulov07, kuznprb08,
kuznphysica13}.

The voltage $V_{rec}(\Phi/\Phi_{0})$ appears due to the difference
in critical currents for positive and negative alternating current
half-waves. It was found \cite{karpii07, nikulov07} that the
critical currents of opposite polarities
$I_{c+}(\Phi/\Phi_{0}+\phi_{l})$ and
$I_{c-}(\Phi/\Phi_{0}-\phi_{l})$ differ not in amplitude, but in
phase shift relative to the zero of the flux in different
directions by the phase difference $\phi_{l}=d \Phi/\Phi_{0}$. The
mutual shift of these critical currents relative to each other
$2\phi_{l}$ can reach 0.5. Any significant mutual phase shift
relative to each other of critical currents of opposite polarities
cannot be obtained from the known geometric and physical
parameters of circularly-asymmetric aluminum structures
\cite{dubjetplet03, karpii07, nikulov07, kuznprb08,
kuznphysica13}. The reason for the appearance of this phase shift
has not yet been found. We proposed a model for the appearance of
this shift, considering a circularly-asymmetric ring as an
asymmetric quantum superconducting interferometer \cite{barone,
clarke}. In this model, which will be presented elsewhere, the
additional phase shift of the critical currents will be nonzero
only if the critical current densities $j_{c1}(T)$ and $j_{c2}(T)$
are not the same in the wide and narrow arms of the
interferometer. For a given geometry of circularly-asymmetric
rings, this is possible only in a situation where the critical
temperatures of the wide and narrow arms of the interferometer
$T_{cw}$ and $T_{cn}$, respectively, differ.

In this work, in order to confirm the assumption about the
difference between $T_{cw}$ and $T_{cn}$, and, consequently, the
difference between $j_{c1}(T)$ and $j_{c2}(T)$ of the wide and
narrow arms of the interferometer, we fabricated wide and narrow
superconducting quasi-one-dimensional aluminum structures with
dimensions close to the typical dimensions of a wide and narrow
aluminum semirings. We measured the critical temperatures and
plotted critical currents as functions of $T$ in wide and narrow
superconducting aluminum structures with different measurement
circuits in zero magnetic field.

\section{Results and Discussion}

\begin{figure}
\begin{center}
\includegraphics[width=1\linewidth]{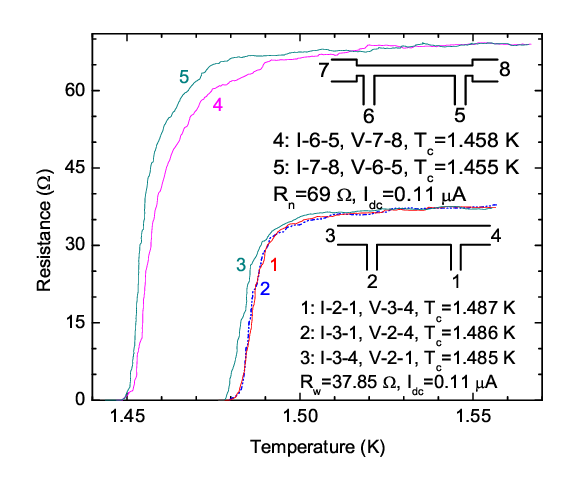}
\caption{\label{f1} (Color online) Lines 1, 2 (dash-dotted), and 3
- resistive $R_{w}(T)$ transitions of the wide structure (bottom
inset) at $I_{dc}=0.11$ $\mu$A. Lines 1, 2, and 3 correspond to
the measurement circuits: ($I$-2-1, $V$-3-4), ($I$-3-1, $V$-2-4),
and ($I$-3-4, $V$-2-1), respectively. Lines 4 and 5 are $R_{n}(T)$
transitions of the narrow structure (inset at the top) at
$I_{dc}=0.11$ $\mu$A. Lines 4 and 5 correspond to the measurement
circuits: ($I$-6-5, $V$-7-8) and ($I$-7-8, $V$-6-5),
respectively.}
\end{center}
\end{figure}

To prove the statement about different critical temperatures of
wide $T_{cw}$ and narrow $T_{cn}$ semirings of a
circularly-asymmetric aluminum interferometer, the narrow and wide
structures (upper and lower insets, Fig. \ref{f1}) were fabricated
on a single silicon chip in one cycle by thermal deposition of an
aluminum film with a thickness of $d=19$ nm using the lift-off
process of electron beam lithography. The widths of the narrow and
wide parts of both structures are approximately the same and equal
to $w_{n}=0.27$ and $w_{w}=0.48$ $\mu$m, respectively. The total
length of both narrow and wide quasi-one-dimensional wires
constituting the structures reached 60 $\mu$m. Long wire length
minimizes influence of wide current and potential contacts of the
structure.

In this work, low voltage electrical signals versus temperature
and applied dc were measured in a room shielded from high
frequency electromagnetic interference. In order to reduce the
influence of low-frequency mains interference and high-frequency
noise on the structures under study, we used, in mainly analog
devices, trying to exclude digital devices. In addition, homemade
low-frequency current generator and dc amplifier powered by
galvanic batteries were used. To minimize the effect of noise on
the structure, one-kilo-ohm resistances were placed on the chip
holder with the structure and connected in series to each current
(potential) contact of the structure.

We measured resistive N-S $R_{w}(T)$ transitions (lines 1, 2, 3 in
Fig. \ref{f1}, below) of  the wide structure and $R_{n}(T)$
transitions (lines 4, 5 in Fig. \ref{f1}, above) of  the narrow
structure with different connection of current $I$ and potential
$V$ wires. The measurement circuits are shown in Fig. \ref{f1}.
Functions $R_{w}(T)$ and $R_{n}(T)$ are recorded in short sections
of the same length 6.69 $\mu$m at an applied dc $I_{dc}=0.11$
$\mu$A. Note that the total length of the current-carrying parts
of both structures for all measurement circuits was approximately
60 $\mu$m. The resistances in the normal state of the narrow and
wide wires are $R_{n}=69$ and $R_{w}=37.85$ $\Omega$,
respectively.

We have determined the critical temperatures $T_{cl}=T_{cl}(0)$,
$T_{c}=T_{c}(0.5)$, and $T_{ch}(0.96)$, corresponding to three
levels of the N-S transition of the wide structure
$R_{w}(T)/R_{w}=0$, 0.5, and 0.96, respectively. The critical
temperatures of the wide structure for different measurement
circuits are equal for line 1 ($I$-2-1, $V$-3-4) - $T_{cl}=1.480$,
$T_{c}=1.487$, $T_{ch}(0.96)=1.521$ K, for line 2 ($I$-3-1,
$V$-2-4) - $T_{cl}=1.480$, $T_{c}=1.486$, $T_{ch}(0.96)=1.522$ K,
for line 3 ($I$-3-4, $V$-2-1) - $T_{cl}=1.478$, $T_{c}=1.485$,
$T_{ch}(0.96)=1.506$ K.

The critical temperatures of the narrow structure are also
determined by the three levels of the N-S transition. The critical
temperatures for the narrow structure for different measurement
circuits are equal for line 4 ($I$-6-5, $V$-7-8) - $T_{cl}=1.449$,
$T_{c}=1.458$, $T_{ch}(0.96)=1.500$ K, for line 5 ($I$-7-8,
$V$-6-5) - $T_{cl}=1.448$, $T_{c}=1.455$, $T_{ch}(0.96)=1.480$ K.

It can be seen that the critical temperatures of the wide and
narrow structures depend on the measurement circuit. Thus, the
critical temperatures of wide and narrow structures found in the
middle of N-S transitions are 1.487 and 1.458 K, in the case when
the measurement circuits ($I$-2-1, $V$-3-4) and ($I$-6-5, $V$-7-8)
are used for wide and narrow structures, respectively (lines 1 and
4, Fig. \ref{f1}). In this case, the voltage is recorded on both
structures using wide wires.

Critical temperatures of the wide and narrow structures found in
the middle of the N-S transitions, are equal to 1.485 and 1.455 K,
in the case when the wide and narrow structures are measured
according to the measurement circuits ($I$-3-4, $V$-2-1) and
($I$-7-8, $V$-6-5), respectively (lines 3 and 5, Fig. \ref{f1}).
In the case, the voltage is measured on both structures using
narrow wires.

We have found that the critical temperature of the narrow
structure is less than the critical temperature of the wide
structure. We believe that critical temperature of narrow wires of
the wide structure is close to the critical temperature 1.455 K of
the narrow structure, recorded according to the measurement
circuit: $I$-7-8, $V$-6-5 (line 5, Fig. \ref{f1}). Similarly, the
critical temperature of wide wires of the narrow structure is
close to the critical temperature of 1.487 K of the wide
structure, recorded according to the measurement circuit: $I$-2-1,
$V$-3-4 (line 1, Fig. \ref{f1}). Thus, at $1.455<T<1.487$ K, both
wide and narrow structures, taken together with current and
potential wires, are heterogeneous normal-superconducting (N-S)
structures. This hybridity of structures can lead to new effects.
We believe that the true critical temperatures (without taking
into account the influence of the proximity effect \cite{schmidt})
have values less than 1.455 K for a narrow wire and values greater
than 1.487 K, for a wide wire.

We briefly explain why critical temperatures depend on the
measurement circuit. The critical temperatures of wide and narrow
structures are higher when the voltage is measured on these
structures using wide wires with a higher critical temperature
$T_{cw}$ than the critical temperature $T_{cn}$ of a narrow wire.
The high density of superconducting electrons $n_{s}$ in parts of
wide wires that do not carry current leads, due to the proximity
effect, to an effective increase in $n_{s}$ and the critical
temperature of the structure.

So, we experimentally found that the difference between the
critical temperatures of wide and narrow aluminum structures
consisting of quasi-one-dimensional superconducting wires of
different widths reaches a value greater than 30\,mK when the wide
and narrow wires of both structures have the same thickness $d=19$
nm and widths equal to $w_{w}=0.48$ and $w_{n}=0.27$ $\mu$m,
respectively.

Earlier, it was experimentally established \cite{chubov} that
critical temperature $T_{c}$ of the thin aluminum film is much
higher than the critical temperature $T_{cb}=1.194$ K of the bulk
superconductor. In \cite{borisenko}, the critical temperature of
narrow sections of submicron width in a superconducting thin-film
aluminum structure of variable width reached 5.8 K. It was found
in \cite{chubov} that the relative increase in the critical
temperature $dT_{c}/T_{cb} \propto 1/d$ and practically does not
depend on the width of $w$ for wide aluminum films with a width of
0.5-1.5 mm. It should be expected that in a quasi-one-dimensional
superconducting aluminum wire with a width satisfying the
condition $d<w<2\xi(T)$, the critical temperature will slightly
increase with decreasing $w$. However, for our aluminum structures
of the same thickness $d$, the critical temperature of the
narrower structure is slightly lower than the critical temperature
of the wider structure.

Here, we propose a mechanism to clarify such a difference in the
critical temperatures of narrow and wide aluminum structures of
the same thickness. The special feature to fabricate our
quasi-one-dimensional superconducting aluminum structures leads to
contamination of the longitudinal boundaries of structures. Dirty
boundaries can contain magnetic atoms or vacancies that electrons
can land on, creating uncompensated electron spin. As a result,
the presence of these magnetic atoms and vacancies causes the
destruction of superconducting pairs and a decrease in the
effective critical temperature of the wire. In our case, the
depairing effect of dirty boundaries is stronger in a narrower
structure than in a wider structure. Therefore, the critical
temperature of the narrow structure is lower than the critical
temperature of the wide structure.

The resistances per square of the narrow and wide structures are
$R_{n}^{sqr}=2.8$ $\Omega$ and $R_{w}^{sqr}=2.7$ $\Omega$,
respectively. Resistivities of the narrow and wide structures
$\rho^{n}_{n}=5.3 \times 10^{-8}$ $\Omega$ m and $\rho^{w}_{n}=5.2
\times 10^{-8}$ $\Omega$ m, respectively, are found from the
expression $\rho=R^{sqr}d$. From the refined theoretical
expression for aluminum $\rho l_{el}=5.1 \times 10^{-16}$ $\Omega$
m$^{2}$ \cite{gershenson}, we obtain the electron mean free paths
in narrow and wide structures $l_{eln}=9.6$ nm and $l_{elw}=9.9$
nm, respectively.

Structures are in a dirty limit since $l_{el}<<\xi_{0}=1.6$
$\mu$m. Near $T_{c}$, the Ginzburg-Landau coherence length is
$\xi(T)=\xi(0)(1-T/T_{c})^{-1/2}$, where
$\xi(0)=0.85(l_{el}\xi_{0})^{1/2}$ is the coherence length at
$T=0$\,K \cite{schmidt}. The temperature-dependent penetration
depth of the magnetic field is
$\lambda_{GL}(T)=\lambda(0)(1-T/T_{c})^{-1/2}$, where
$\lambda(0)=0.615\lambda_{L}(\xi_{0}/l_{el})^{1/2}$ is the depth
penetration of the field at $T=0$\,K, $\lambda_{L}=16$ nm is the
London penetration depth for aluminum \cite{schmidt}. Since
$l_{eln}$ and $l_{elw}$ are narrow and wide structures are close,
we get close coherence lengths $\xi_{n}(0) \approx \xi_{w}(0)
\approx 0.11$ $\mu$m and the field penetration depth $\lambda_{n}
(0) \approx \lambda_{w}(0) \approx 0.13$ $\mu$m. Indexes $n$ and
$w$ refer to narrow and wide structures, respectively.
Temperature-dependent density of the Ginzburg-Landau depairing
current is equal to $j_{GL}(T)=j_{GL}(0)(1-T/T_{c})^{3/2}$
\cite{schmidt}. The critical current density at $T=0$ is
$j_{GL}^{n}(0) \approx j_{GL}^{w}(0) \approx 6.0 \times 10^{10}$
A/m\,$^{2}$ for narrow and wide wires. Critical depairing currents
at $T=0$ for narrow ($w_{n}=0.27$ $\mu$m, $d=19$ nm) and wide
wires ($w_{w}=0.48$ $\mu$m, $d=19$ nm) are equal to
$I_{GL}^{n}(0)=303$ $\mu$A and $I_{GL}^{w}(0)=550$ $\mu$A,
respectively.

In order to prove the statement about different critical current
densities of wide $j_{c1}(T)$ and narrow $j_{c2}(T)$ semirings of
a circularly-asymmetric aluminum interferometer, we measured the
voltage $V(I)$ as a function of the applied dc $I$ in wide and
narrow structures at $T$ slightly below $T_{c}$ in zero magnetic
field with different measurement circuits. Note that the voltage
was recorded on a short section of a rather long superconducting
quasi-one-dimensional aluminum wire. $V(I)$ curves have thermal
hysteresis depending on the direction of the current sweep. Using
the $V(I)$ curves, we plotted the temperature dependences of the
switching and retrapping critical currents in wide and narrow
structures. The switching current switches the structure from a
superconducting state to a resistive state. The retrapping current
is the current at which structure returns from the resistive state
into the superconducting state.

For a more accurate comparison of the measured critical switching
current with the theory over a wider temperature range, we used
the phenomenological expression for temperature-dependent critical
current density of the dirty quasi-one-dimensional superconducting
wire $j_{c}(T)=
(j_{KL}(0)/4)(1-(T/T_{c})^{2})(1-(T/T_{c})^{4})^{1/2}$, where
$j_{KL}(0)=((8\pi^{2}\sqrt{2\pi})/(21\zeta(3)e))\sqrt{(kT_{c})^{3}/(\hbar
v_{F}\rho_{n} \rho_{n}l_{el})}$ is critical current density in the
dirty limit at $T=0$, calculated in \cite{romijn} within the
framework of the Kupriyanov-Lukichev theory \cite{kupriyanov}. The
expression for $j_{c}(T)$ is obtained using the phenomenological
expressions $\lambda(T)=\lambda(0)/\sqrt{1-(T/T_{c})^{4}}$ and
$H_{c}(T)=H_{c}(0)(1-(T/T_{c})^{2})$ \cite{schmidt, tinkham}.

The critical current density at $T=0$\,K, calculated within the
framework of the Kupriyanov-Lukichev theory, is equal to
$j_{KL}^{n}(0)=7.2 \times 10^{10}$ A/m\,$^{2}$ for a narrow wire
($\rho^{n}_{n}=5.3 \times 10^{-8}$ $\Omega$ m, $T_{c}=1.455$ K).
For wide wire ($\rho^{w}_{n}=5.2 \times 10^{-8}$ $\Omega$ m,
$T_{cw}=1.486$ K) this critical current density is equal to
$j_{KL}^{w}(0)=7.6 \times 10^{10}$ A/m\,$^{2}$.

The values of the Kupriyanov-Lukichev critical currents at $T=0$
for a narrow wire ($w_{n}=0.27$ $\mu$m, $d=19$ nm) and a wide wire
($w_{w}=0.48$ $\mu$m, $d=19$ nm) are equal to $I_{KL}^{n}(0)=371$
$\mu$A and $I_{KL}^{w}(0)=690$ $\mu$A, respectively. Then
$I_{KL}^{n}(0)/4=93$ $\mu$A and $I_{KL}^{w}(0)/4=173$ $\mu$A. The
Kupriyanov-Lukichev critical currents differ from the
corresponding Ginzburg-Landau critical currents, equal to
$I_{GL}^{n}(0)=303$ $\mu$A and $I_{GL}^{w}(0)=550$ $\mu$A.

\begin{figure}
\begin{center}
\includegraphics[width=1\linewidth]{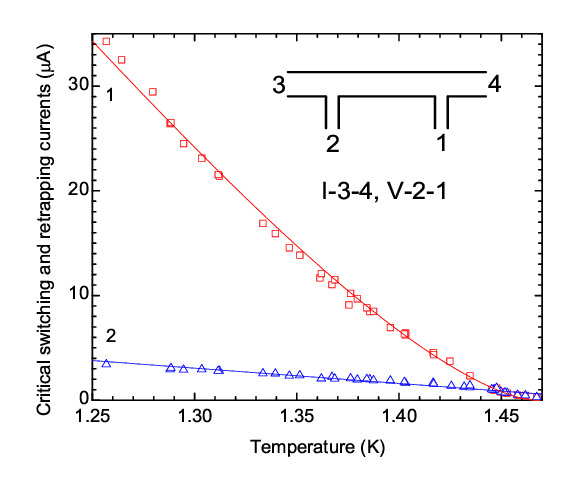}
\caption{\label{f2} (Color online) Squares and triangles are
critical switching and retrapping currents as functions of the
temperature, respectively, recorded in the wide structure
according to the measurement circuit: $I$-3-4, $V$-2-1. Lines 1
and 2 are adjustable theoretical functions for the switching
current $I_{c1}(T)$ and the retrapping current $I_{ret2}(T)$,
respectively. Inset is the sketch of the structure.}
\end{center}
\end{figure}

Figure \ref{f2} shows the critical switching current (squares) and
the retrapping current (triangles) as functions of $T$ measured
over several cycles in wide structure ($I$-3-4, $V$-2-1) and
corresponding adjustable lines 1 and 2. Inset of Fig. \ref{f2}
represents the sketch of the structure. Line 1 shows the
approximation of the experimental switching current at
$T=1.25-1.466$ K using the expression
$I_{c1}(T)=I_{cf1}(0)(1-(T/T_{cf1})^{2})(1-(T/T_{cf1})^{4})^{1/2}$,
where the adjustable critical current $I_{cf1}(0)=183$ $\mu$A and
the adjustable critical temperature $T_{cf1}=1.466$ K. The value
of $I_{cf1}(0)$ is close to the expected theoretical value
$I_{KL}^{w}(0)/4=173$ $\mu$A. We measured, for the first time,
that the adjustable critical temperature $T_{cf1}$, determined
from the temperature dependence of the switching current
$I_{c1}(T)$, less than the critical temperature $T_{c}=1.486$ K,
found in the middle of the N-S transition of the structure.
Moreover, the adjustable critical temperature $T_{cf1}$ less than
the critical temperature $T_{cl}=1.478$ K, corresponding to the
bottom of the N-S transition, by 12 mK. We experimentally checked
that the adjustable critical temperature $T_{cf1}$ and the
critical temperature $T_{c}$ coincide for the control
quasi-one-dimensional aluminum structure with the same width of
the current and potential wires.

Line 2 represents the linear fit of the experimental retrapping
current at $T=1.25-1.445$ K by the expression
$I_{ret2}(T)=I_{retf2}(0)(1-T/T_{cf2})$, where the adjustable
critical current $I_{retf2}(0)=22$ $\mu$A and the adjustable
critical temperature $T_{cf2}=1.510$ K. The adjustable critical
temperature $T_{cf2}=1.510$ K is close to the critical temperature
$T_{ch}(0.96)=1.506$ K, that corresponds to the top of the N-S
transition of the wide structure. We believe that the adjustable
critical temperature $T_{cf2}$ is close to the true critical
temperature of a wide wire without taking into account the effect
of narrow parts of the structure. We assume that the linear
dependence of the retrapping current at $T=1.25-1.445$ K is
determined by the Joule (or quasiparticle) overheating of the
structure.

At $T=1.445-1.466$ K, the measured switching and retrapping
critical currents coincide in experimental error (no hysteresis)
and lie on line 1, that is below the linear function
$I_{ret2}(T)$.

\begin{figure}
\begin{center}
\includegraphics[width=1\linewidth]{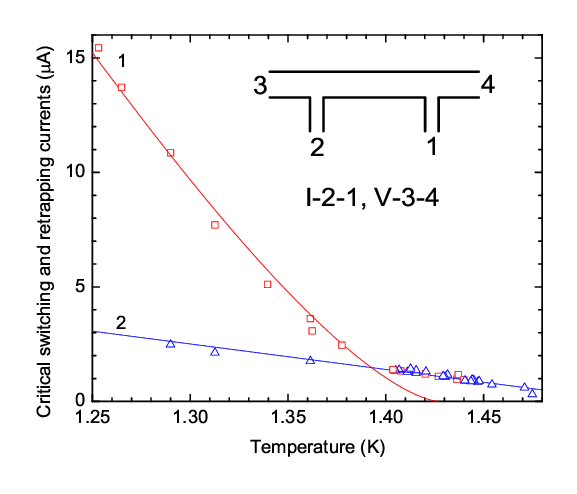}
\caption{\label{f3} (Color online) Squares and triangles -
switching and  retrapping currents as functions of the
temperature, respectively, recorded in a wide structure according
to the measurement circuit: $I$-2-1, $V$-3-4. Line 1 is the
adjustable curve $I_{c1}(T)$ for the experimental switching
current in the range $T=1.25-1.4$ K. Line 2 is the adjustable
curve $I_{ret2}(T)$ for the switching current in the interval
$T=1.404-1.475$ K and for the retrapping current in the interval
$T=1.25-1.475$ K. Inset is a sketch of the structure.}
\end{center}
\end{figure}

Figure \ref{f3} shows the critical switching (squares) and
retrapping (triangles) currents as functions of  the temperature,
recorded over several cycles in the wide structure according to
the measurement circuit ($I$-2-1, $V$-3-4) and adjustable lines 1
and 2. The inset in Fig. \ref{f3} demonstrates the structure
sketch. Note that in the case of the measurement circuit $I$-2-1,
$V$-3-4, the switching current is mainly determined by the
narrowest part of the structure. We found, for the first time,
that the measured switching current as a function of  the
temperature is approximated by two theoretical curves (lines 1 and
2). We draw attention that for this structure, recorded by another
measurement circuit: $I$-3-4, $V$-2-1 (Fig. \ref{f2}) and for the
control quasi-one-dimensional aluminum structure with one width,
the experimental switching current is approximated by a single
nonlinear temperature dependence.

Line 1 is the fit of the experimental switching current at
$T=1.25-1.4$ K by the expression
$I_{c1}(T)=I_{cf1}(0)(1-(T/T_{cf1})^{2})(1-(T/T_{cf1})^{4})^{1/2}$,
here the adjustable critical current $I_{cf1}(0)=102$ $\mu$A and
the adjustable critical temperature $T_{cf1}=1.427$ K. The value
of $I_{cf1}(0)$ is close to the expected theoretical value for a
narrow structure $I_{KL}^{n}(0)/4=93$ $\mu$A. We found that the
same as for the switching current (Fig. \ref{f2}), the adjustable
critical temperature $T_{cf1}$ and the critical temperature
$T_{c}=1.487$ K, determined from the middle N-S transition of the
structure, do not match. In addition, the value $T_{cf1}$ is less
than the critical temperature $T_{cl}=1.480$ K, corresponding to
the bottom of the N-S transition of the wide structure by 53 mK.
We assume that the value $T_{cf1}=1.427$ K is close to the true
critical temperature of a narrow wire in the structure (without
taking into account the proximity effect). In the interval
$T=1.404-1.475$ K, the experimental switching current, which
coincides with the retrapping current, is approximated by line 2,
which is given by the linear dependence
$I_{ret2}(T)=I_{retf2}(0)(1-T/T_{cf2})$, where the adjustable
critical current $I_{retf2}(0)=17$ $\mu$A and the adjustable
critical temperature $T_{cf2}=1.525$ K. The adjustable critical
temperature $T_{cf2}=1.525$ K is close to the critical temperature
$T_{ch}(0.96)=1.521$ K, which corresponds to the top of the N-S
transition of the wide structure. We believe that the value of
$T_{cf2}$ is close to the true critical temperature of a wide
wire, without taking into account the effect of narrow parts of
the structure. Thus, the difference in the assumed true critical
temperatures of the wide and narrow wires is equal to
$T_{cf2}-T_{cf1}=98$ mK.

We believe that at $T=1.404-1.475$ K, a wide structure measured
according to the measurement circuit $I$-2-1, $V$-3-4 (Fig.
\ref{f3}), is a hybrid structure, consisting of two SNS junctions
formed at the points of connection of narrow current wires. We
assume that in the interval $T=1.404-1.475$ K, the function
$I_{ret2}(T)$ coincides with the critical current of the Josephson
structure $I_{J}(T)$. Near $T_{c}$,  $I_{J}(T)=\pi
\Delta^{2}(T)/(4ekTR_{J})$, where
$\Delta(T)=1.74\Delta(0)(1-T/T_{c})^{1/2}$ is a
temperature-dependent energy superconducting gap in zero magnetic
field, $\Delta(0)=1.764kT_{c}$ is a gap at $T=0$, $R_{J}$ is the
Josephson resistance \cite{likharev}. The expression
$I_{J}(T)=I_{J}(0)(1-T/T{c})$ can be written, where $I_{J}(0)$ is
the Josephson critical current at $T=0$ K. In our case,
$I_{J}(0)=I_{retf2}(0)=17$ $\mu$A, $T_{c}=T_{cf2}=1.525$ K. In
this case, the value of $I_{J}(0)$ corresponds to the Josephson
resistance $R_{J}=57.2$ $\Omega$, that is close to the resistance
of the narrow structure $R_{n}(T)=69$ $\Omega$.

Line 2 is the fit of the experimental retrapping current at
$T=1.25-1.475$ K using the same expression
$I_{ret2}(T)=I_{retf2}(0)(1-T/T_{cf2})$, as for the switching
current at $T=1.404-1.475$ K. The adjustable critical current
$I_{retf2}(0)=17$ $\mu$A and the adjustable critical temperature
$T_{cf2}=1.525$ K. In the interval $T=1.25-1.4$ K, the whole
structure is superconducting and linear dependence of the
retrapping current $I_{ret2}(T)$ is determined by another reason,
namely, Joule (or quasiparticle) overheating of the structure.

\begin{figure}
\begin{center}
\includegraphics[width=1\linewidth]{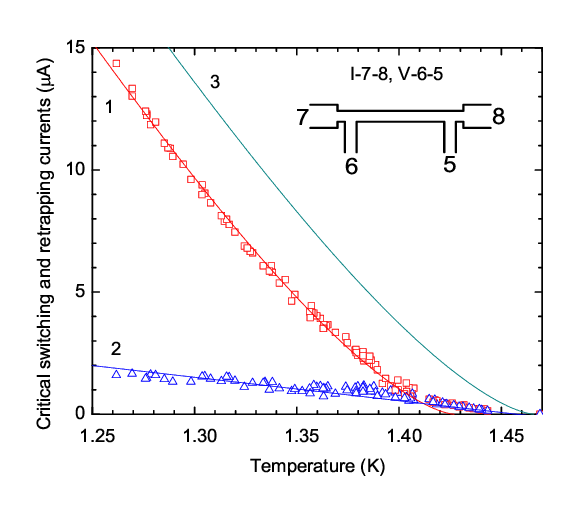}
\caption{\label{f4} (Color online) Squares and triangles are the
switching and retrapping currents as functions of  the
temperature, respectively, recorded in a narrow structure
according to the measurement circuit: $I$-7-8, $V$-6-5. Line 1 is
the adjustable curve $I_{c1}(T)$ for the experimental switching
current in the range $T=1.25-1.41$ K. Line 2 is the adjustable
curve $I_{ret2}(T)$ for the switching current at $T=1.415-1.444$ K
and for the retrapping current at $T=1.25-1.45$ K. Line 3 is the
expected switching current of a narrow structure if critical
temperatures of narrow and wide wires making up a narrow and wide
structure, are the same. Inset: a sketch of the structure.}
\end{center}
\end{figure}

Figure \ref{f4} shows the critical switching (squares) and
retrapping (triangles) currents as functions of  the temperature,
measured in several cycles in a narrow structure ($I$-7-8,
$V$-6-5) and adjustable lines 1 and 2. The inset in Fig. \ref{f4}
represents a sketch of the structure. We found that for this
narrow structure, as well as for the wide structure measured
according to the measurement circuit $I$-2-1, $V$-3-4 (Fig.
\ref{f3}), the experimental switching current as a function of $T$
is approximated with using two theoretical curves (lines 1 and 2).

Line 1 is the fit of the measured switching current at
$T=1.25-1.41$ K using the expression
$I_{c1}(T)=I_{cf1}(0)(1-(T/T_{cf1})^{2})(1-(T/T_{cf1})^{4})^{1/2}$,
here the adjustable critical current $I_{cf1}(0)=102$ $\mu$A and
the adjustable critical temperature $T_{cf1}=1.427$ K. The value
of $I_{cf1}(0)$ is close to the expected theoretical value
$I_{KL}^{n}(0)/4=93$ $\mu$A. Note that for this switching current,
as well as for switching currents (Figs. \ref{f2}, \ref{f3}), the
adjustable critical temperature $T_{cf1}$ and the critical
temperature $T_{c}=1.455$ K found in the middle of the N-S
transition of the structure are different. Moreover, the value
$T_{cf1}=1.427$ K is less than the critical temperature
$T_{cl}=1.448$ K, corresponding to the bottom of the N-S
transition of the narrow structure at 21 mK. We believe that the
value of $T_{cf1}$ is close to the true critical temperature of a
narrow wire of the structure (without taking into account the
proximity effect). Expressions for line 1 (Fig. \ref{f3}) and for
line 1 (Fig. \ref{f4}) coincide.

In the interval $T=1.415-1.444$ K, the measured switching current
coincides with the retrapping current, and is given by line 2,
that is described by the linear dependence
$I_{ret2}(T)=I_{retf2}(0)(1-T/T_{cf2})$, where the adjustable
critical current $I_{retf2}(0)=14$ $\mu$A and the adjustable
critical temperature $T_{cf2}=1.457$ K. The adjustable critical
temperature $T_{cf2}$ is the effective critical temperature of a
narrow wire with allowance for the effect of wide parts of the
structure. The adjustable critical temperature $T_{cf2}$ is less
than the critical temperature close to the top of the N-S
transition $T_{ch}(0.96)=1.480$ K.

We believe that at $T=1.415-1.444$ K, the narrow structure, taken
together with the wide current wires, behaves like a hybrid SNS
structure. Note that at $T=1.404-1.475$ K, a wide structure,
measured according to the measurement circuit $I$-2-1, $V$-3-4
(Fig. \ref{f3}), also represents a hybrid SNS structure. We
believe that at $T=1.415-1.444$ K, the function $I_{ret2}(T)$
coincides with the critical current of the Josephson structure
$I_{J}(T)=I_{J}(0)(1-T/T{c})$. In our case,
$I_{J}(0)=I_{retf2}(0)=14$ $\mu$A, $T_{c}=T_{cf2}=1.457$ K. The
critical Josephson current at zero temperature $I_{J}(0)$
corresponds to the Josephson resistance $R_{J}=66.35$ $\Omega$,
which is close to the structure resistance $R_{n}(T)=69$ $\Omega$.

Line 2 is the fit of measured retrapping current at $T=1.25-1.45$
K by the same expression $I_{ret2}(T)=I_{retf2}(0)(1-T/T_{cf2})$,
as for the switching current at $T=1.415-1.444$ K. The adjustable
critical current $I_{retf2}(0)=14$ $\mu$A and the adjustable
critical temperature $T_{cf2}=1.457$ K. At $T=1.25-1.41$ K, the
entire structure is superconducting and linear dependence of the
retrapping current $I_{ret2}(T)$ is due to the Joule (or
quasiparticle) overheating of the structure.

We found that the switching and retrapping critical currents as
functions of the temperature of the investigated narrow structure
measured using other measurement circuit ($I$-6-5, $V$-7-8) are
approximated with the same expressions $I_{c1}(T)$ and
$I_{ret2}(T)$. This data are not listed here.

Line 3 (Fig. \ref{f4}) shows the expected switching critical
current of a narrow structure if the critical temperatures of the
narrow and wide parts of the structures are equal. The expression
for line 3 is the product of the ratio $w_{n}/w_{w}$ and the
expression for adjustable switching critical current of wide
structure (line 1, Fig. \ref{f2}).

Thus, we measured that for the same thickness $d=19$, the critical
current density in a narrow quasi-one-dimensional superconducting
aluminum structure is less than the critical current density in a
wide aluminum structure.

In addition, our results indicate that the temperature-dependent
critical current in cases where the voltage is measured over a
short section of long quasi-one-dimensional superconducting
aluminum structures is a non-local value. This current depends on
the temperature-dependent superconducting order parameter and the
critical temperature of the sections of current and potential
wires located outside this short section.

Additionally, to confirm the values of the true critical
temperatures of wide and narrow aluminum wires, we measured
control wide and narrow aluminum structures, consisting of current
and potential wires of the same width. These structures were
fabricated on a single chip and had a thickness and widths, close
to the thickness and widths of wide and narrow wires that make up
the wide and narrow structures studied here. The distances between
potential contacts were the same for both control structures and
were close to the distances between potential contacts of the
structures presented here.

We found the critical temperatures determined by the middle of the
N-S transition for the control wide and narrow structures are
close to the assumed true critical temperatures of wide and narrow
wires making up the investigated wide and narrow structures. In
addition, we found that the critical switching currents of the
control wide and narrow structures in the entire investigated
temperature range near $T_{c}$ are well approximated by similar
theoretical expressions
$I_{c}(T)=I_{cf}(0)(1-(T/T_{cf})^{2})(1-(T/T_{cf})^{4})^{1/2}$,
where the adjustable critical currents $I_{cf}(0)$ are close to
the corresponding theoretical values $I_{KL}(0)/4$ and the
adjustable critical temperatures $T_{cf}$ are close to the
corresponding critical temperatures $T_{c}$ found in the middle of
the N-S transition. Note that the linear temperature dependence of
the switching current at $T$ very close to $T_{c}$, inherent in
hybrid S-N-S structure, is not observed on the control wide and
narrow structures.

\section{Conclusion}

Thus, we have shown experimentally that the critical temperatures
$T_{cw}$ and $T_{cn}$ of wide and narrow quasi-one-dimensional
superconducting aluminum structures with typical dimensions close
to the sizes of a wide and narrow semirings of the
circularly-asymmetric interferometer are not the same.

For the first time, we found that a decrease in the width $w$ of
quasi-one-dimensional superconducting aluminum wires of the same
thickness $d$ leads to decrease in the critical temperature of the
wire. At a time when a slight increase in the critical temperature
is expected with decreasing $w$. We believe that the narrower the
wire the higher the influence of depairing centers located on
dirty longitudinal boundaries of the wire and, therefore, the
lower the critical temperature.

In order to prove that the difference in the densities of the
critical switching current $j_{c1}(T)$ and $j_{c2}(T)$ of wide and
narrow aluminum structures with sizes close to those of wide and
narrow arms of a circularly-asymmetric aluminum interferometer, we
measured the switching and retrapping critical currents in these
structures. The values of critical currents were obtained from the
hysteresis $V(I)$ curves recorded at an applied dc in zero
magnetic field at temperatures close to $T_{c}$. The $V(I)$ curves
are recorded using different electrical measurement circuits.

We found that at $T$ lower $T_{cf1}$, the switching current as a
function of $T$ in wide and narrow structures for different
measurement circuits is well described by the expression
$I_{c1}(T)=I_{cf1}(0)(1-(T/T_{cf1})^{2})(1-(T/T_{cf1})^{4})^{1/2}$,
where $I_{cf1}(0)$ and $T_{cf1}$ are the adjustable switching
current at $T=0$  and the adjustable critical temperature. The
adjustable switching current $I_{cf1}(0)$ is close to the value
$I_{KL}(0)/4$ calculated within the framework of the
Kupriyanov-Lukichev theory \cite{kupriyanov}. First, it was found
that the adjustable critical temperature $T_{cf1}$ is lower than
the critical temperature $T_{cl}$ corresponding to the bottom of
the N-S transition and is significantly lower than the critical
temperature $T_{c}$ determined from the middle of the N-S
transition of the structure. Then, as for the control
quasi-one-dimensional aluminum structure with the same width of
current and potential wires, the critical temperatures $T_{cf1}$
and $T_{c}$ are the same.

We found that the switching current for a wide structure, recorded
according to the measurement circuit: $I$-3-4, $V$-2-1 (Fig.
\ref{f2}), is approximated in the entire experimental temperature
range by one dependence $I_{c1}(T)$, as well as for the control
quasi-one-dimensional aluminum wire of the same width. It was
found for the first time for cases (Figs. \ref{f3}, \ref{f4}) that
the switching temperature-dependent current of wide and narrow
structures is described by two functions. At lower temperatures,
this switching current is given by the $I_{c1}(T)$ function. At
higher temperatures, Josephson SNS junctions are formed in
structures. In the range of higher temperature, the switching
current equal to the retrapping current is described by the linear
function $I_{ret2}(T)=I_{retf2}(0)(1-T/T_{cf2})$, which coincides
with the Josephson critical current $I_{J}(T)=I_{J}(0)(1-T/T{c})$
, where the adjustable critical current $I_{J}(0)=I_{retf2}(0)$
and the adjustable critical temperature $T_{c}=T_{cf2}$. It was
found that the adjustable critical temperature $T_{cf2}$ is close
to the critical temperature $T_{ch}(0.96)$, which corresponds to
the top of the N-S transition in the structures.

For the entire experimental temperature range, except for the case
(Fig. \ref{f2}), the retrapping current as a function of $T$ in
wide and narrow structures with different connections of current
and potential wires is approximated by the same expression
$I_{ret2}(T)=I_{retf2}(0)(1-T/T_{cf2})$ with the same adjustable
critical current $I_{retf2}(0)$ and the adjustable critical
temperature $T_{cf2}$ as for the switching current at higher
temperatures. The linear dependence of the retrapping current
$I_{ret2}(T)$ is due to the Joule (or quasiparticle) overheating
of the structure at lower temperatures and the linear function
$I_{J}(T)$ at higher temperatures.

In addition, we plotted a theoretical temperature dependence of
the switching critical current of a narrow structure with one
width and a critical temperature equal to the critical temperature
of a wide structure. It turned out that the calculated switching
current is much higher than the experimental switching current of
the narrow structure under study.

Thus, for the first time, we have shown experimentally for
thin-film quasi-one-dimensional superconducting aluminum
structures of the same thickness $d=19$ nm, that the critical
current density in a narrow structure under study is lower than
the critical current density in a wide structure under study.

In addition, we found that the critical current of
quasi-one-dimensional superconducting aluminum structures is a
nonlocal quantity.

This finding that critical temperatures and critical current
density of wide and narrow quasi-one-dimensional superconducting
aluminum wires of the same thickness differ, can help clarify the
effects observed in thin-film aluminum wires and structures
consisting of wires with different widths \cite{kuznjetplet16,
dubjetplet03, karpii07, nikulov07, kuznprb08, kuznphysica13,
kuznjetplet19}.

In addition, the results of this work allows explaining the
negative local and nonlocal voltage (resistance) in a
quasi-one-dimensional superconducting aluminum wire of variable
width \cite{kuznphysica20} and a mysterious phase shift in the
magnetic field of critical currents of different polarity in
opposite directions in circularly-asymmetric aluminum structures
\cite{karpii07, nikulov07}.

\section{Acknowledgments}

We are grateful to A. Firsov for the fabrication of aluminum
structures and M. Skvortsov for fruitful discussions. This work
was carried out and financially supported within the framework of
STATE TASK No. 075-00355-21-00.



\end{document}